\documentclass[twocolumn,aps]{revtex4-1}
\usepackage[]{natbib}
\usepackage{graphicx}  
\usepackage{dcolumn}   
\usepackage{bm}        
\usepackage{verbatim}   
\usepackage{graphicx}
\usepackage{epstopdf}
\usepackage{footnote}
\usepackage{epsfig}
\usepackage{subfigure}
\usepackage{amssymb}
\usepackage{amsmath,bm}
\usepackage{xcolor}
\usepackage{float}
\usepackage{subfigure}
\usepackage[most]{tcolorbox}
\usepackage{soul}
\usepackage{color}
\usepackage[colorlinks=true,linkcolor=blue,allcolors=blue]{hyperref}%
\usepackage[framemethod=TikZ]{mdframed}
\usepackage{lipsum}
\mdfdefinestyle{MyFrame}{%
	linecolor=blue,
	outerlinewidth=2pt,
	roundcorner=20pt,
	innertopmargin=\baselineskip,
	innerbottommargin=\baselineskip,
	innerrightmargin=20pt,
	innerleftmargin=20pt,
	backgroundcolor=gray!50!white}

\begin{document}
	
	
	\headsep = 40pt
	\title{Space-Time Medium Functions as a Perfect Antenna-Mixer-Amplifier Transceiver}
	\author{Sajjad Taravati and George V. Eleftheriades}
	\affiliation{The Edward S. Rogers Sr. Department of Electrical and Computer Engineering, University of Toronto, Toronto, Ontario M5S 3H7, Canada\\
		Email: sajjad.taravati@utoronto.ca}

\begin{abstract}
	We show that a space-time-varying medium can function as a front-end transceiver, i.e., an antenna-mixer-amplifier. Such a unique functionality is endowed by space-time surface waves associated with complex space-time wave vectors in a subluminal space-time medium. The proposed structure introduces pure frequency up- and down-conversions and with very weak undesired time harmonics. In contrast to other recently proposed space-time mixers, a large frequency up-/down conversion ratio, associated with gain is achievable. Furthermore, as the structure does not operate based on progressive energy transition between the space-time modulation and the incident wave, it possesses a subwavelength thickness (metasurface). Such a multi-functional, highly efficient and compact medium is expected to find various applications in modern wireless telecommunication systems.
\end{abstract}

\maketitle

\section{Introduction}

Reception and transmission of electromagnetic waves is the essence of wireless telecommunications~\cite{vannucci1995wireless,saunders2007antennas}. Such a task requires various functions, such as power amplification, frequency conversion, and wave radiation. These functions have been conventionally performed by electronic components and electromagnetic structures, such as transistor-based amplifiers, semiconductor-based diode mixers, and resonance-based antennas. The ever increasing demand for versatile wireless telecommunication systems has led to a substantial need in multi-functional components and compact integrated circuits. In general, each natural medium may introduce a single function, e.g., a resonator may operate as an antenna at a specific frequency. To contrast this conventional approach, a single medium exhibiting several functions concurrently could lead to a disruptive evolution in telecommunication technology.    

Recently, space-time-modulated media have attracted a surge of scientific interest thanks to their capability in multifunctional operations, e.g., mixer-duplexer-antenna~\cite{Taravati_LWA_2017}, unidirectional beam splitters~\cite{Taravati_Kishk_PRB_2018}, nonreciprocal filters~\cite{alvarez2019coupling,wu2019isolating}, and signal coding metagratings~\cite{taravati_PRApp_2019}. In addition, a large number of versatile and high efficiency electromagnetic systems have been recently reported based on the unique properties of space-time modulation, including space-time metasurfaces for advanced wave engineering and extraordinary control over electromagnetic waves~\cite{Fan_APL_2016,Fan_mats_2017,Salary_2018,salary2019dynamically,Taravati_Kishk_TAP_2019,zang2019nonreciprocal_metas,inampudi2019rigorous,elnaggar2019generalized,wang2018photonic,Grbic2019serrodyne,Taravati_Kishk_MicMag_2019,ptitcyn2019time,sedehtopological,du2019simulation,wang2019multifunctional,taravati2019full,li2020time,sedeh2020topological}, nonreciprocal platforms~\cite{Taravati_thesis,wentz1966nonreciprocal,Taravati_PRB_2017,Taravati_PRB_SB_2017,Taravati_PRAp_2018,oudich2019space,chegnizadeh2020non}, frequency converters~\cite{Taravati_PRB_Mixer_2018,Grbic2019serrodyne}, and time-modulated antennas~\cite{shanks1961new,Alu_PNAS_2016,ramaccia2018nonreciprocity,taravati2018space,salary2019nonreciprocal,zang2019nonreciprocal}. Such strong capability of space-time-modulated media is due to their unique interactions with the incident field~\cite{Taravati_PRB_2017,li2019nonreciprocal,correas2018magnetic,liu2018huygens,du2019simulation,elnaggar2019modelling,taravati2019_mix_ant}. 

\begin{figure*}
	\subfigure[]{\label{fig:3D_slab_dl} 
		\includegraphics[width=1\columnwidth]{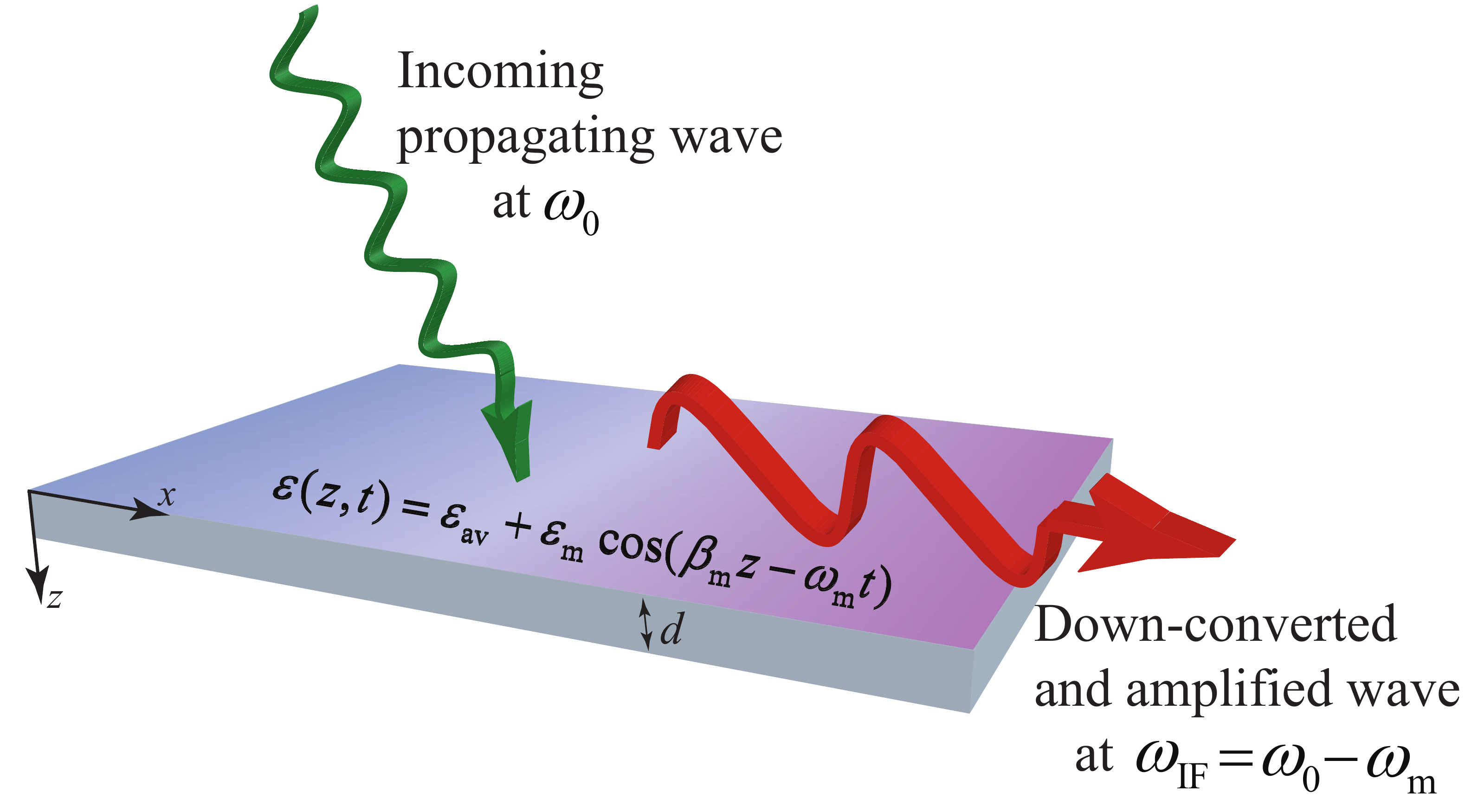}} 
	\subfigure[]{\label{fig:3D_slab_ul} 
		\includegraphics[width=1\columnwidth]{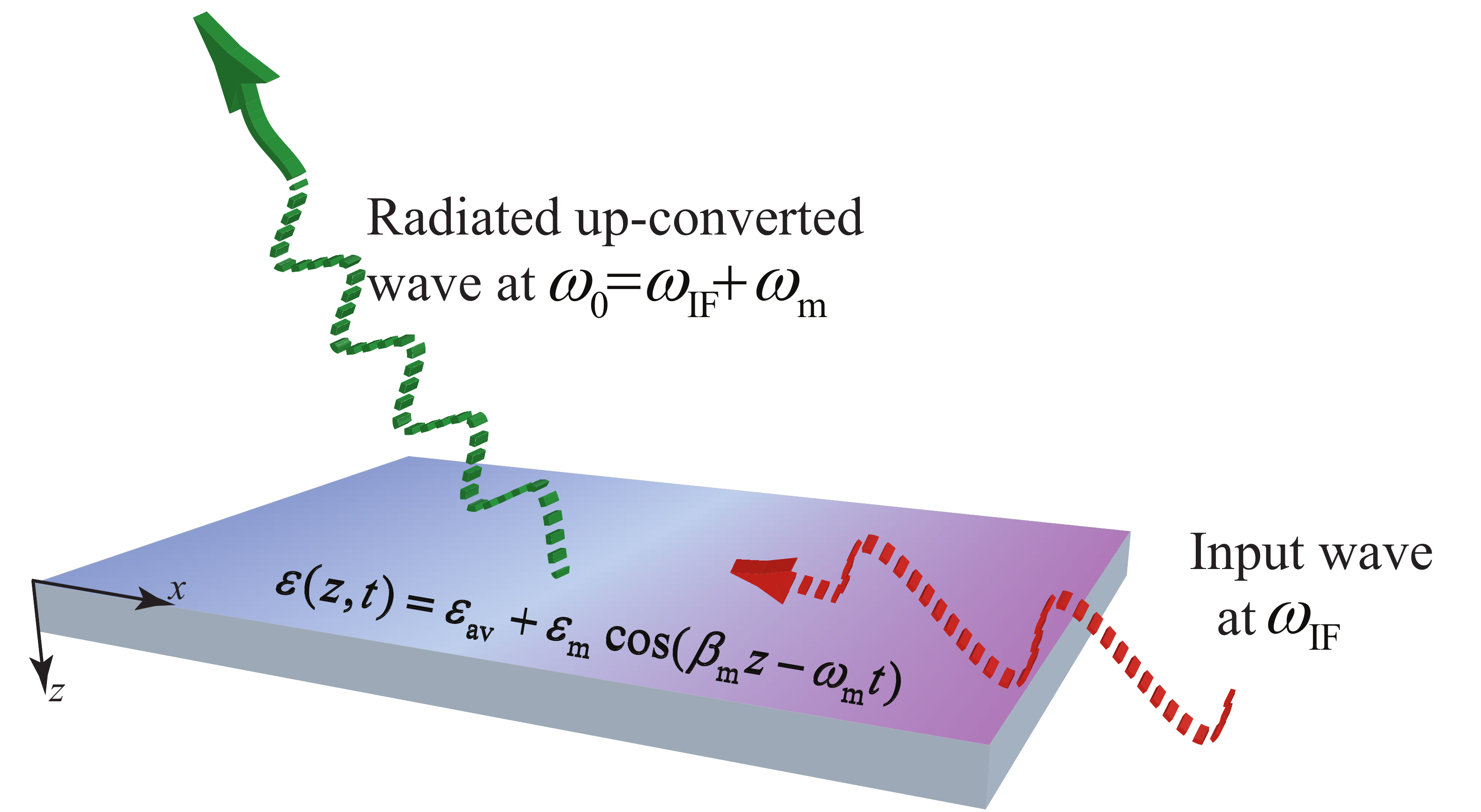}}
	\subfigure[]{\label{fig:symb}
		\includegraphics[width=1.3\columnwidth]{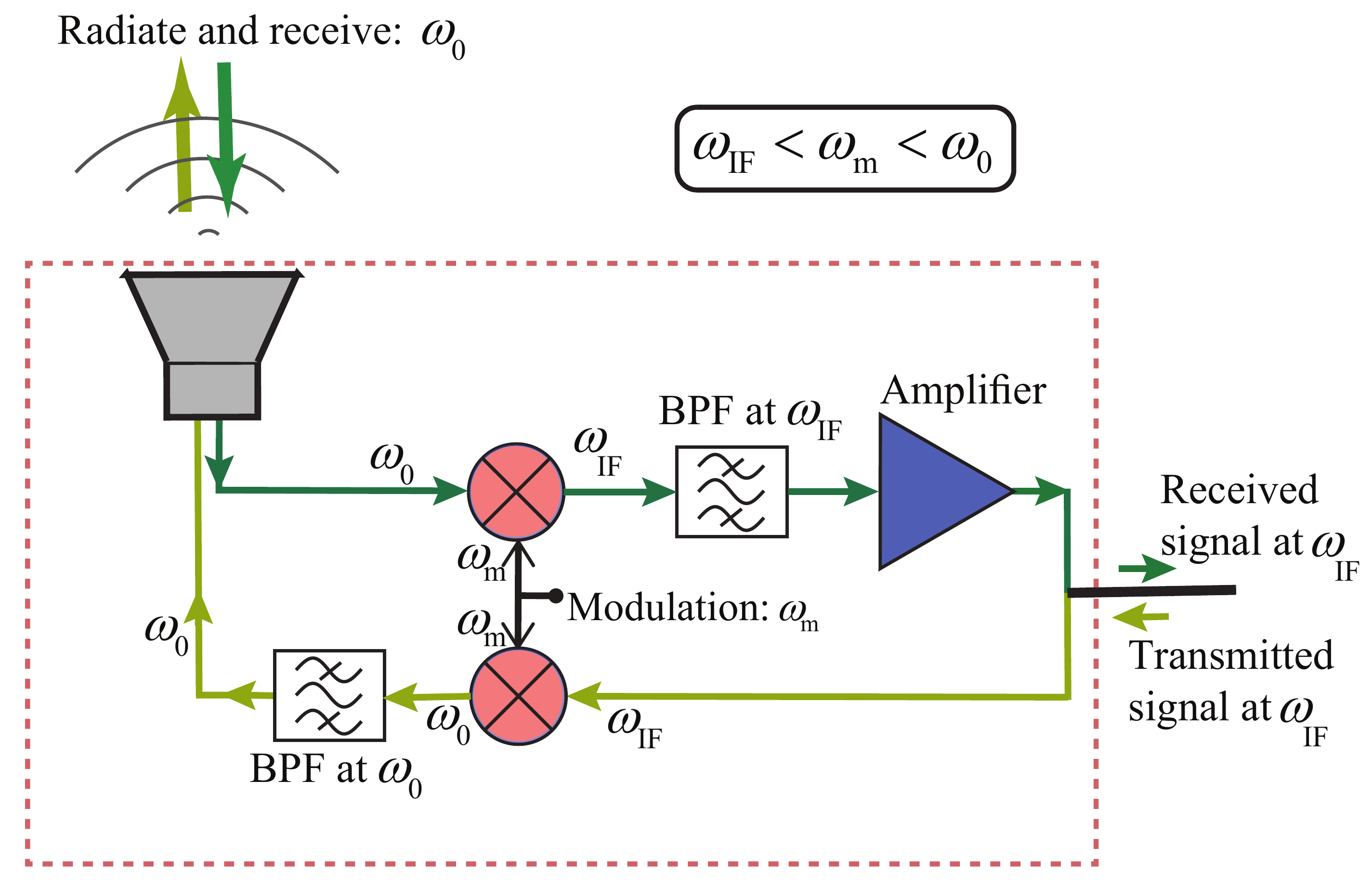}} 
	\caption{Functionality of the antenna-mixer-amplifier space-time surface wave medium. (a) Down-link (reception) state, where a strong transition between a space-wave at $\omega_0$ to a space-time surface wave at $\omega_\text{IF}=\omega_0-\omega_\text{m}$ occurs. (b) Up-link (transmission) state, where a transition between a space-time surface wave at $\omega_\text{IF}$ to a space-wave at $\omega_\text{0}=\omega_\text{IF}+\omega_\text{m}$ occurs. (c) Circuital representation of the combined antenna-mixer-amplifier space-wave medium in (a) and (b).
	} 
	\label{Fig:slab_concept}
\end{figure*}

This paper reveals that a space-time medium can function as a full transceiver front-end, that is, an antenna-mixer-amplifier-filter system. Specific related contributions of this study are as follows.

\begin{itemize}
	\item We show that the space-time medium operates as an antenna per se.  Such an interesting functionality of the space-time medium is endowed by space-time surface waves. Other recently proposed space-time antenna systems are formed by integration of the space-time-modulated medium with an antenna~\cite{Taravati_APS_2015,Alu_PNAS_2016,Taravati_LWA_2017}, and hence suffer from a number of drawbacks, i.e., requiring long structures, low efficiency and narrow-band operation. 
	
	\item The proposed antenna-mixer-amplifier introduces large frequency up- and -down conversion ratios. This is very practical, because in real-scenario wireless telecommunication systems, a large frequency conversion is required, i.e., a frequency conversion from a microwave/millimeter-wave frequency to an intermediate frequency in receivers. In contrast, recently proposed space-time frequency converters suffer from very low frequency conversion ratios (up-/down-converted frequency is very closed to the input frequency)~\cite{Taravati_APS_2015,Alu_PNAS_2016,Taravati_LWA_2017,Grbic2019serrodyne}.

	\item Recently proposed space-time systems operate based on a progressive coupling between the incident wave and the space-time modulation~\cite{Fan_NPH_2009,Fan_PRL_109_2012,Taravati_LWA_2017,Chamanara_PRB_2017,Taravati_PRB_Mixer_2018}, and hence a thick (compared to the wavelength) slab is required. In contrast, the proposed antenna-mixer-amplifier does not work based on progressive coupling between the incident wave and the space-time modulation and hence is formed by a thin (sub-wavelength) slab. As a result, the proposed structure is classified among metasurface categories and is compatible with the compactness requirements of modern wireless telecommunication systems.
		
	\item The proposed medium inherently operates as a band-pass filter, i.e., a pure frequency down- and up-conversions occurs so that the undesired time harmonics are highly suppressed. Such a property is governed by the subluminal operation of space-time modulation.	
	
	\item In contrast to conventional mixers that introduce a significant conversion loss, here a conversion gain may be achieved in the down-link or up-link. In addition, the radiation from the antenna-mixer-amplifier is very directive.
\end{itemize}

The paper is organized as follows. Section~\ref{sec:oper} presents the operation principle of the antenna-mixer-amplifier. Then, Sec.~\ref{sec:theo} investigates the theoretical aspects of the work, and provides an analysis of the wave propagation in the antenna-mixer-amplifier space-time medium. Next, Sec.~\ref{sec:disgn} presents the design procedure of the structure, and gives an approximate closed-form solution to calculate the thickness and receive/transmit power gain of the structure.
Section~\ref{sec:Res} gives the time and frequency domain numerical simulation results for the designed antenna-mixer-amplifier.
Finally, Sec.~\ref{sec:conc} concludes the paper.

\section{Operation Principle}\label{sec:oper}

Consider the antenna-mixer-amplifier slab in Fig.~\ref{fig:3D_slab_dl}, with the thickness of $d$ and a spatiotemporally-periodic electric permittivity as
\begin{equation}\label{eqa:perm}
\epsilon(z,t)=\epsilon_\text{av} +\epsilon_\text{m} \cos(\beta_\text{m} z-\omega_\text{m} t),
\end{equation}
with $\epsilon_\text{av}$ being the average electric permittivity of the background medium, and $\epsilon_\text{m}$ representing the modulation amplitude. In Eq.~\eqref{eqa:perm}, $\omega_\text{m}$ denotes the temporal frequency of the modulation, and $\beta_\text{m}$ is the spatial frequency of the modulation represented by
\begin{equation}\label{eqa:vm}
\beta_\text{m}= \dfrac{\omega_\text{m}}{v_\text{m}}=\dfrac{\omega_\text{m}}{\varGamma v_\text{b}},
\end{equation}
where $v_\text{m}$ and $v_\text{b}$ are the phase velocity of the modulation and the background medium, respectively, and $\varGamma=v_\text{m}/v_\text{b}$ is the space-time velocity ratio~\cite{Taravati_PRB_2017}.

The slab in Fig.~\ref{fig:3D_slab_dl} is obliquely illuminated by a $y$-polarized incident electric filed under an angle of incidence of $\theta_{\text{i}}$, as
\begin{equation}\label{eqa:Ei}
\mathbf{E}_\text{i} (x,z,t)= \mathbf{\hat{y}} E_\text{i} e^{i\left( k_x x +k_{z} z- \omega_0 t  \right) },
\end{equation}
where $E_\text{i}$ is the amplitude of the incident wave, and $\omega_0$ and $k_\text{0}=\sqrt{k_x^2 +k_{z}^2}=\sqrt{(k_0 \sin(\theta_{\text{i}}))^2 +(k_0 \cos(\theta_{\text{i}}))^2}$ are respectively temporal and spatial frequencies of the incident wave. Here, $k_0=\omega_0/c$, with $c$ being the velocity of light in vacuum.

In the receiving state (down-link) in Fig.~\ref{fig:3D_slab_dl}, strong transition between a space-wave, with temporal frequency $\omega_0$, to a space-time surface wave, with temporal frequency $\omega_\text{IF}=\omega_0-\omega_\text{m}$, occurs. In the transmission state (up-link) in Fig.~\ref{fig:3D_slab_ul} a transition between a space-time surface wave at $\omega_\text{IF}$ to a space-wave at $\omega_\text{0}=\omega_\text{IF}+\omega_\text{m}$ occurs.

Figure~\ref{fig:symb} shows an equivalent circuit representation of the antenna-mixer-amplifier in Figs.~\ref{fig:3D_slab_dl} and~\ref{fig:3D_slab_ul}. Thanks to the unique properties of space-time-modulated media, that will be described later in this study, such a medium introduces the functionality of a highly directive antenna, a pure frequency up-converter, a pure frequency down-converter, up-link and down-link filters, and a down-link amplifier. Such a rich functionality has not been experienced in other media unless several components and media are integrated, as shown in Fig.~\ref{fig:symb}. However, here we show that a single medium can offer such versatile and useful operation.

As the slab is periodic in both space and time, the spatial and temporal frequencies of the space-time harmonics inside the structure are governed by the momentum conservation law, i.e.,
\begin{subequations}
	\begin{equation}\label{eqa:gr_eq01}
	\gamma_{z,n}=\beta_{z,0}+ n \beta_\text{m}+i\alpha_{z,n},
	\end{equation} 
	and the energy conservation law, i.e.,
	\begin{equation}\label{eqa:gr_eq14}
	\omega_n=\omega_0+ n \omega_\text{m}.
	\end{equation} 
\end{subequations}

The scattering angles of the space-time harmonics in regions 1 and 3 (reflection to the top and transmission to the bottom of the medium, respectively) may be determined by the Helmholtz relations, i.e., $k_{0}^2 \sin^2(\theta_{\text{i}})+k_{0,n}^2 \cos^2(\theta_{\text{r}n})=k_{0,n}^2$ and $k_{0}^2 \sin^2(\theta_{\text{i}})+k_{0,n}^2 \cos^2(\theta_{\text{t}n})=k_{0,n}^2$, with $\theta_{\text{r}n}$ and $\theta_{\text{t}n}$ being the reflection and transmission angles of the $n^\text{th}$ space-time harmonics, reading~\cite{taravati_PRApp_2019}
\begin{equation}\label{eqa:refl_trans_angl}
\theta_{\text{r}n}=\theta_{\text{t},n}=\sin^{-1} \left(\frac{\sin(\theta_\text{i})}{1+n\omega_\text{m} /\omega_0} \right),
\end{equation}
where $k_{0n}=\omega_n/c=(\omega_0+n \omega_\text{m})/c$. Therefore, given the common tangential wavenumber, $k_x=k_0\sin(\theta_\text{i})$ in all the regions, the reflection and transmission angles for the $n^\text{th}$ space-time harmonic are equal. Equation~\eqref{eqa:refl_trans_angl} shows that the harmonics in the $n$-interval $[\omega_0(\sin\theta_\text{i}-1)/\omega_\text{m},+\infty]$ are scattered at different angles ranging from $0$ to $\pi/2$ through $\theta_\text{i}$ for $n=0$. However, the space-time harmonics outside of this interval represent imaginary $k_{z,n}$ and hence propagate as space-time surface waves along the two boundaries of the slab at $z=0$ and $z=d$. 

The scattering angles of the space-time harmonics inside the space-time-modulated medium read
\begin{equation}\label{eqa:mod_angl}
\theta_{n}=\tan^{-1} \left(\frac{k_{x}}{\beta_{z,n}} \right)=\tan^{-1} \left(\frac{k_{0} \sin(\theta_{\text{i}})}{\beta_{z,0} + n \beta_\text{m} }\right).
\end{equation}

\begin{figure}
		\includegraphics[width=1\columnwidth]{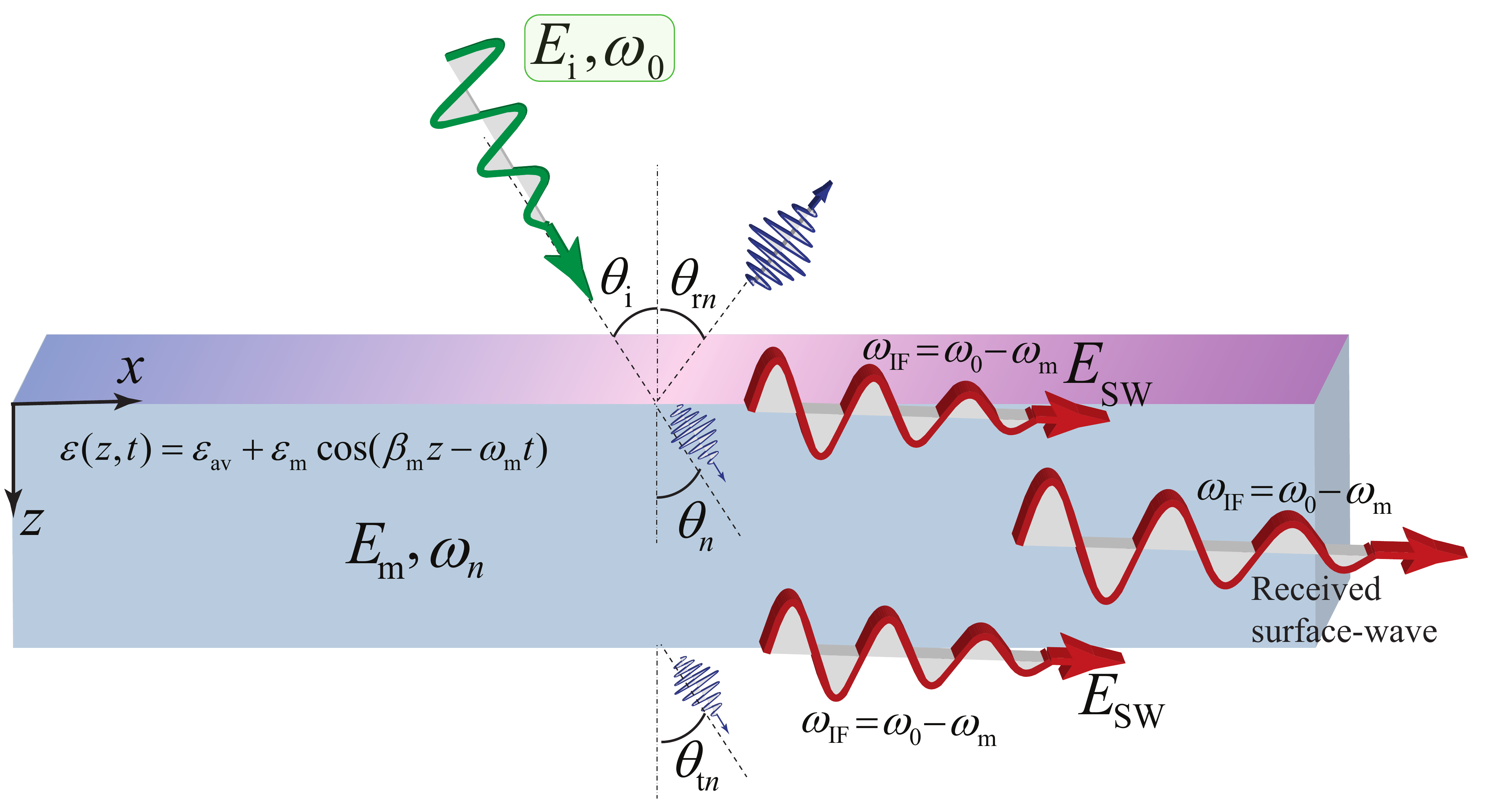}	 	
	\caption{Operation principle of the antenna-mixer-amplifier space-time medium. Transition of the incident space-wave to space-time surface waves at the two boundaries of the medium ($\theta_{\text{r},-1}=\theta_{\text{t},-1}=90^\circ$) as well as inside it ($\theta_{-1}=90^\circ$)).}
	\label{fig:2D_slab} 
\end{figure}

Figure~\ref{fig:2D_slab} shows the structure of the antenna-mixer-amplifier space-time medium. We aim to design the structure such that the first lower space-time harmonic, $n=-1$, outside the medium scatters along the boundary of the medium, i.e., $\theta_{\text{r},-1}=\theta_{\text{t},-1}=90^\circ$. Hence, using Eq.~\eqref{eqa:refl_trans_angl}, the incident angle reads
\begin{equation}\label{eqa:ttt1}
\theta_\text{i}=\sin^{-1} \left(1-\frac{\omega_\text{m}}{\omega_0} \right).
\end{equation}

In addition, to achieving a strong transition to the $n=-1$ harmonic, the scattered $n=-1$ harmonic inside the medium should propagate in parallel to the two space-time surface waves on the two boundaries of the medium at $z=0$ and $z=d$, i.e., $\theta_{n=-1}=90^\circ$. Thus, using Eq.~\eqref{eqa:mod_angl}, we achieve
\begin{equation}\label{eqa:ttt}
\beta_{z,-1}=0. 
\end{equation}

As a result, the $z$ component of the Wave vector inside the medium is purely imaginary, i.e., $\gamma_{z,-1}=i\alpha_{z,-1}$, whereas the incident field wavenumber $k_z$ is purely real.

The space-time-modulated medium presents transition from the fundamental harmonic $n=0$ to a large (theoretically infinite, $-\infty<n<\infty$) number of space-time harmonics. Such a transition is very strong for the luminal space-time modulation, where the space-time modulation velocity is close to the background phase velocity, i.e., $v_\text{m} = v_\text{b}$~\cite{Taravati_PRB_2017,Taravati_PRAp_2018}.

To prevent generation of strong undesired time harmonics, here the space-time-modulated medium operates in the subluminal regime, where $0< v_\text{m}< v_\text{b}$, i.e.,
\begin{equation}\label{eqa:sonic}
0< \varGamma_{\text{subluminal}} <\sqrt{\frac{\epsilon_\text{r}}{\epsilon_\text{av}+\epsilon_\text{m}}}.
\end{equation}

As a result, a pure and precise transition between the fundamental ($n=0$) harmonic and the desired (here $n=-1$) space-time surface wave harmonic can occur.

\section{Theoretical investigation}\label{sec:theo}
\subsection{Analytical solution}\label{sec:anal_sol}

Considering the TE$_{y}$ incident field in Eq.~\eqref{eqa:Ei}, the incident magnetic field reads
\begin{equation}
	\begin{split}
	\mathbf{H}_\text{i}(x,z,t)&= \frac{1}{\eta_1}\left[\mathbf{\hat{k}}_\text{i} \times \mathbf{E}_\text{i}^\text{F} (x,z,t)\right] \\
	&=  \left[-\mathbf{\hat{x}} \cos(\theta_\text{i}) +\mathbf{\hat{z}} \sin(\theta_\text{i}) \right] \dfrac{E_0 }{\eta_1} e^{i\left( k_x x +k_{z} z- \omega _0 t  \right)},
	\end{split}
\end{equation}
\noindent where $\eta_1=\sqrt{\mu_0 /\epsilon_0}$ is the characteristic impedance of free space. Since the space-time-modulated medium is periodic in space and time, the electric and magnetic fields in the slab may be decomposed to space-time Bloch-Floquet harmonics, as
\begin{subequations}
	\begin{equation}
	\mathbf{E}_\text{m}(x,z,t)=\mathbf{\hat{y}}\sum_{n }   \textbf{E}_{n}  e^{ i \gamma_{z,n} z}  e^{  i \left( k_x x-\omega_n t \right)},
	\label{eqa:A-E_mod_field}
	\end{equation}
	and
	\begin{equation}
	\begin{split}
	\mathbf{H}_\text{m}(x,z,t)=
	\dfrac{1}{k} \sum_{n } \textbf{H}_{n}   \left[ -\mathbf{\hat{x}} \beta_{z,n} + \mathbf{\hat{z}} k_x \right]   e^{ i \gamma_{z,n} z} e^{  i \left( k_x x-\omega_n t \right)} .
	\end{split}
	\label{eqa:A-H_mod_field}
	\end{equation}
\end{subequations}

The unknown coefficients in Eqs.~\eqref{eqa:A-E_mod_field} and~\eqref{eqa:A-H_mod_field}, $\gamma_{z,n}=\beta_{z,n}+i\alpha_{z,n}$, $E_{n}$ and $H_{n}$ shall be found through the following procedure. We first construct and solve the corresponding wave equation. Next, we fix the source frequency $\omega_0$ and then find the corresponding discrete $\gamma_{z,n}$ solutions, which form the dispersion diagram of the space-time medium. We next apply the spatial boundary conditions at the edges of the slab, i.e. at $z=0$ and $z=d$, for all the $(\omega_0,\gamma_{z,n})$ states in the dispersion diagram. This provides the unknown field coefficients $E_{n}$ and $H_{n}$ inside the slab, as well as the scattered (reflected and transmitted) fields outside the slab, i.e., $E_\text{R}$ and $E_\text{T}$.

The homogeneous wave equation reads
\begin{equation}
\nabla^2 \mathbf{E}_\text{m}(x,z,t) - \frac{1}{{{c^2}}}\frac{{{\partial ^2} \left[\epsilon (z,t)\mathbf{E}_\text{m}(x,z,t) \right]}}{{\partial {t^2}}}=0.
\label{eqa:wav_eq}
\end{equation}

Inserting~\eqref{eqa:A-E_mod_field} and~\eqref{eqa:perm} into~\eqref{eqa:wav_eq}, and using the Fourier series expansion for the permittivity, results in 
\begin{equation}
E_{n} \left[ \frac{ k_x^2+ (\beta_{z,n} + i \alpha_{z,n})^2  }{\left[(\omega_0+n\omega_\text{m})/c \right]^2 }\right]
-  \sum\limits_{k =  - \infty }^\infty  \tilde{\epsilon}_k E_{n-k}  =0.
\label{eqa:recurs_gen}
\end{equation}
where $\tilde{\epsilon}_k$ are the Fourier series coefficients of the permittivity. Equation~\eqref{eqa:recurs_gen} may be cast to the matrix form as
\begin{equation}
[K]\cdot[E_n]=0,
\label{eqa:matrix_eq}
\end{equation}
where $[K]$ is the $(2N+1)\times(2N+1)$ matrix with elements
\begin{equation}
\begin{split}
K_{nn} &= \left[ \frac{ k_x^2+ (\beta_{z,n} + i \alpha_{z,n})^2  }{\left[(\omega_0+n\omega_\text{m})/c \right]^2 }\right]-  \epsilon_0,   \\
K_{nk} &= -  \tilde{\epsilon}_{n-k},\quad\text{for }n\neq k,
\end{split}
\label{eqa:K_matrix}
\end{equation}
where $2N+1$ is the number of truncated harmonics. Then, the dispersion relation reads
\begin{equation}
\text{det}\left\{[K]\right\}=0,
\label{eqa:det_gen}
\end{equation}

\begin{figure}
		\includegraphics[width=1.0\columnwidth]{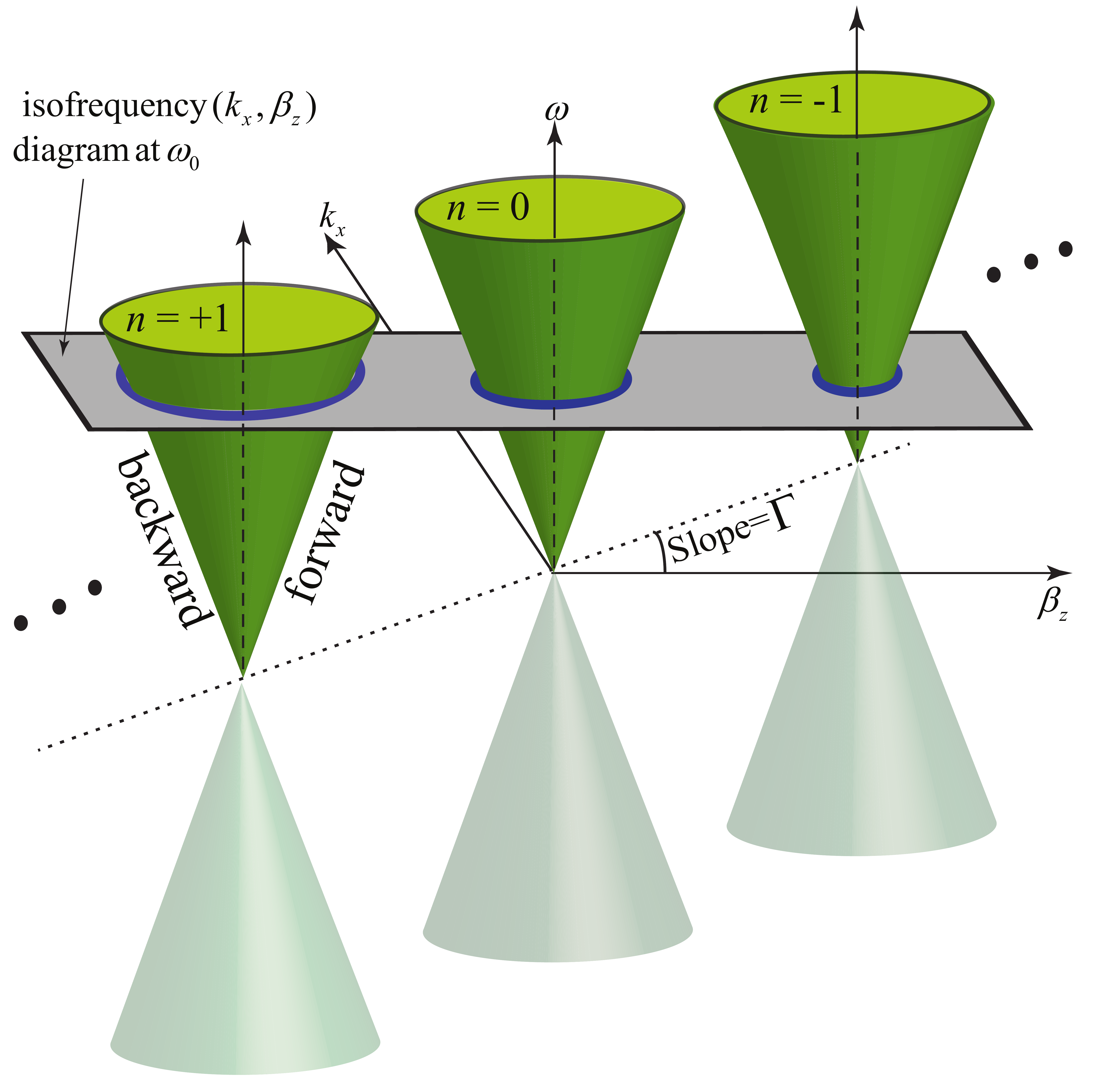}	 	
	\caption{3D-dispersion diagram of the space-time medium, formed by an infinite number of double cones each of which representing a space-time harmonic.}
	\label{fig:3D_dispis} 
\end{figure}

Once the dispersion diagram is formed, for a given incident temporal frequency $\omega_0$, the corresponding wave number, i.e., $\gamma_{z,n}$, inside the slab can be computed. Next, the unknown field amplitudes $E_{n}$ in the slab are found by solving~\eqref{eqa:matrix_eq} after determining the $E_{0}$ satisfying boundary conditions. Figure~\ref{fig:3D_dispis} shows a qualitative 3D dispersion diagram of a subluminal space-time modulated medium, composed of an infinite periodic set of double-cones with
apexes at $\beta_{z,0}=-n\beta_\text{m}$ and $k_x=0$ and the slope $\varGamma$. As the slope $\varGamma$ increases and goes towards unity, the cones get closer to each other and start touching other cones(harmonics). As a result, stronger transition between harmonics occurs and yields transitions of the power from the fundamental harmonic to a large number of harmonics. However, this is not what we aim at. As we are interested in a pure transition to a desired harmonic ($n=-1$), we require a subluminal space-time modulation defined in Eq.~\eqref{eqa:sonic}.

\begin{figure*}
		\includegraphics[width=2\columnwidth]{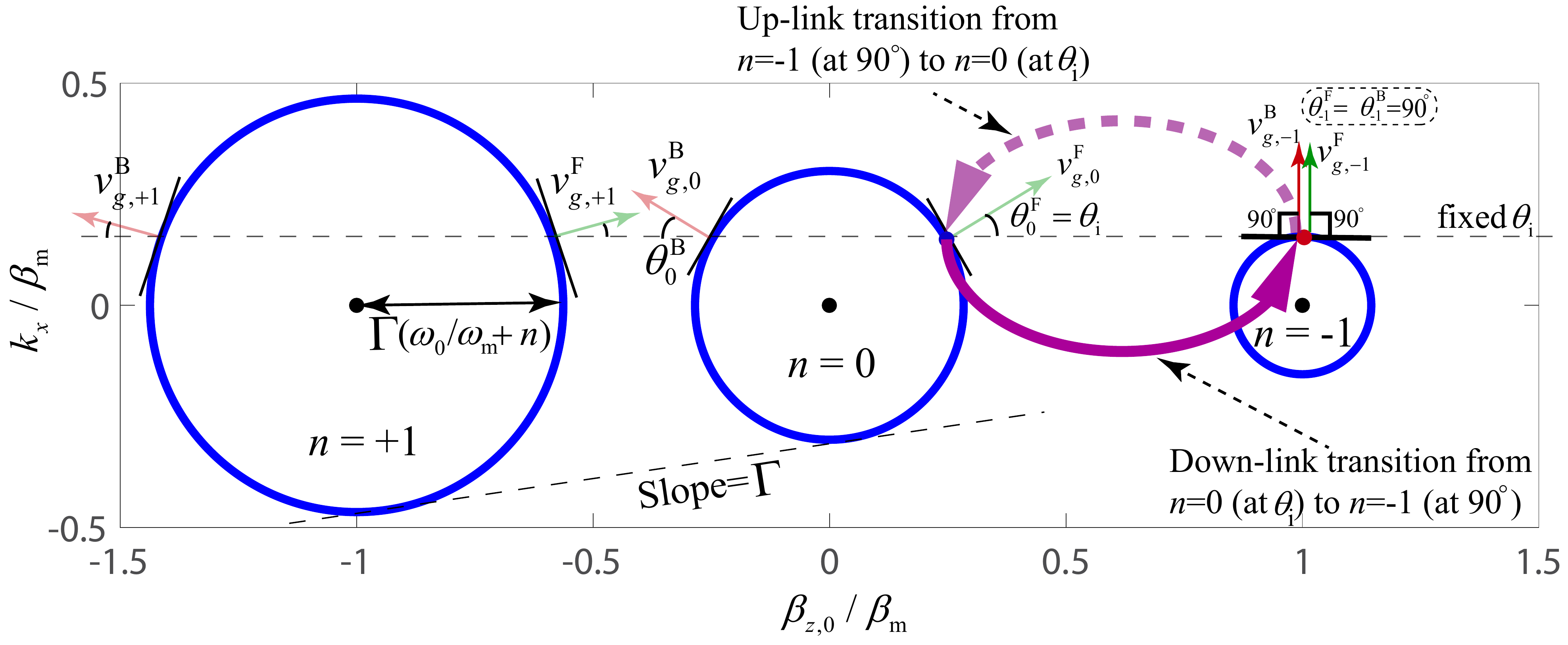}
	\caption{Analytical isofrequency dispersion diagram for $\varGamma=0.2$ and 	$\epsilon_\text{m}=\rightarrow 0$, showing down-link (reception) and up-link (transmission) electromagnetic transitions between the fundamental propagating harmonic ($n=0$) and the space-time surface wave harmonic ($n=-1$). Excitation of the medium at $\theta_\text{i}$ leads to excitation of the $n=-1$ at $90^\circ$.}
	\label{fig:dispiso_1} 
\end{figure*}

Figure~\ref{fig:dispiso_1} plots the analytical isofrequency dispersion diagram, computed using~\eqref{eqa:det_gen}, for the subluminal regime where $\varGamma=0.2$. We have chosen a specific $\theta_\text{i}$, where the $n=-1$ harmonic is excited at a point corresponding to $\beta_{z,-1}=0$ (i.e., $\beta_{z,0}/\beta_\text{m}=+1$). As a consequence, the $n=-1$ harmonic scatters along the $x$-direction, i.e., $\theta_{\text{r},-1}=90^\circ$. Although the real part of the $z$ component of the wavevector of the $n=-1$ harmonic is zero ($\beta_{z,-1}=0$), its imaginary part is greater than zero, and hence $\gamma_{z,-1}=i\alpha_{z,-1}$.

Figure~\ref{fig:dispiso_1} shows that the transition between the forward $n=0$ and $n=-1$ harmonics (highlighted with magenta arrows) occurs for both reception (down-link) and transmission (up-link). This may as well be verified by 
Eq.~\eqref{eqa:mod_angl} as follows. For the down-link, the incident wave ($n=0$) with the incident angle of $\theta_\text{i}^\text{RX}$ and wavenumber of $\beta_{z,0}^\text{RX}=\beta_\text{m}$ makes a transition to the down-converted harmonic ($n=-1$), which is attributed to $\theta_\text{-1}^\text{RX}=90^\circ$, $\beta_{z,-1}^\text{RX}=0$ and $\gamma_{z,-1}^\text{RX}=i\alpha_{z,-1}^\text{RX}$. However, for the up-link (transmission state), the incident wave ($n=0$) with the incident angle of $\theta_\text{i}^\text{TX}=90^\circ$ and wavenumber of $\beta_{z,0}^\text{TX}=0$ makes a transition to the up-converted harmonic ($n=+1$) with $\theta_\text{+1}^\text{TX}=\theta_\text{0}^\text{RX}=\theta_\text{i}^\text{RX}$, which is attributed to $\beta_{z,+1}^\text{TX}=\beta_\text{m}$ and $\gamma_{z,+1}^\text{TX}=\beta_\text{m}$.

\section{Design Procedure}\label{sec:disgn}

\subsection{Complex Dispersion Diagrams}\label{sec:disp}

\begin{figure*}
	\subfigure[]{\label{fig:dispiso_2}
		\includegraphics[width=1\columnwidth]{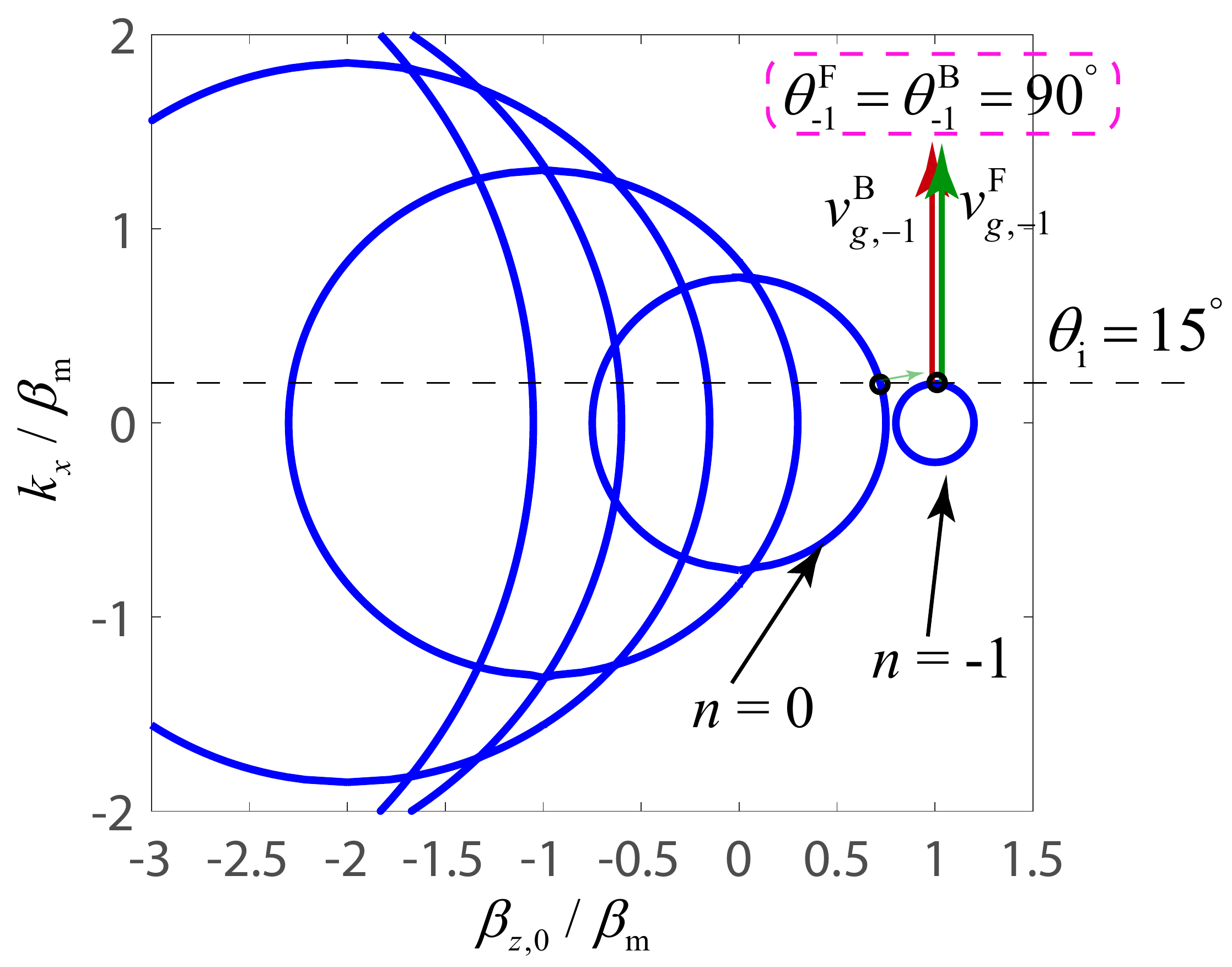}}
	\subfigure[]{\label{fig:dispiso_3}
		\includegraphics[width=1\columnwidth]{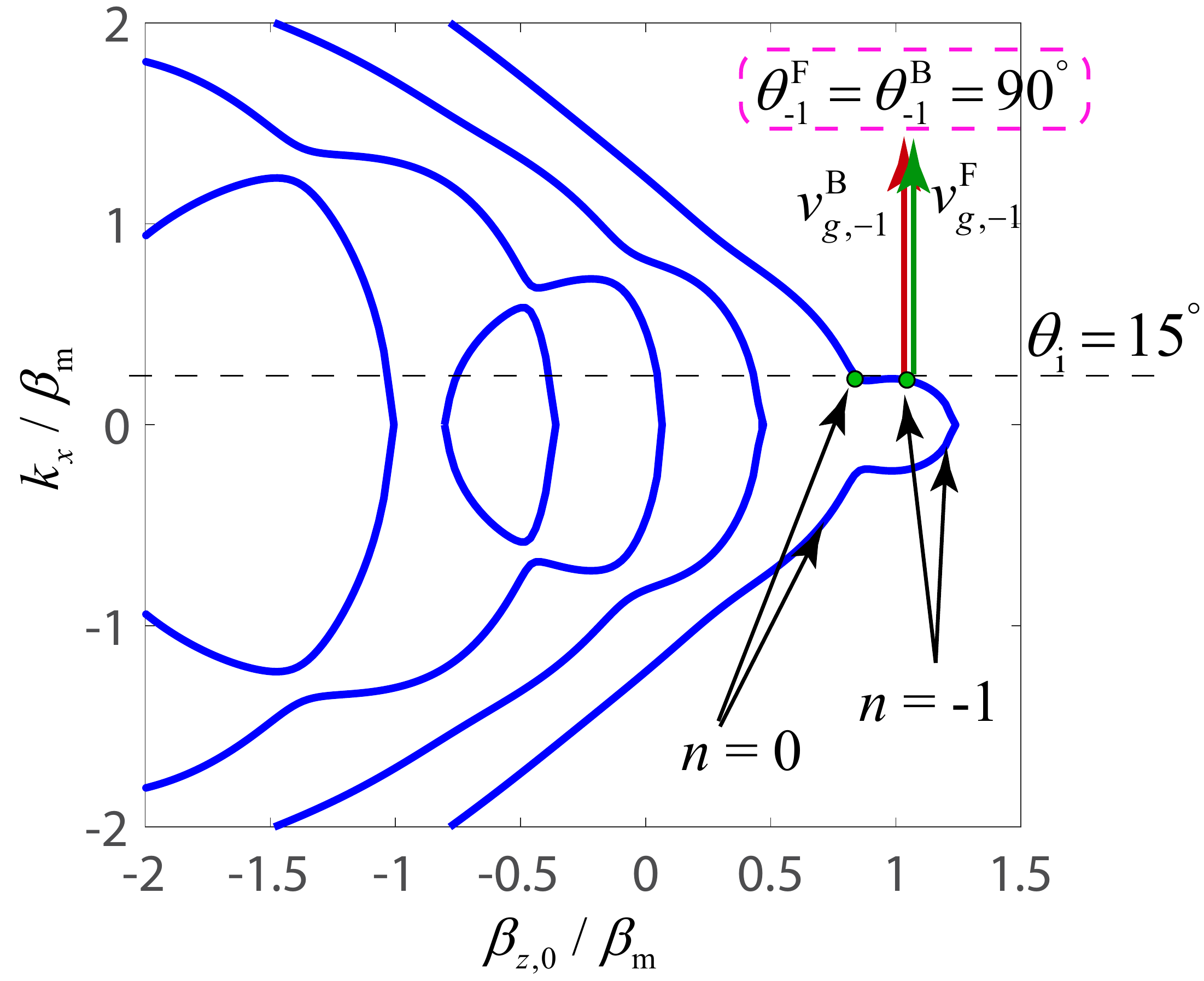}} 	
	\subfigure[]{\label{fig:dispiso_4}	
		\includegraphics[width=0.9\columnwidth]{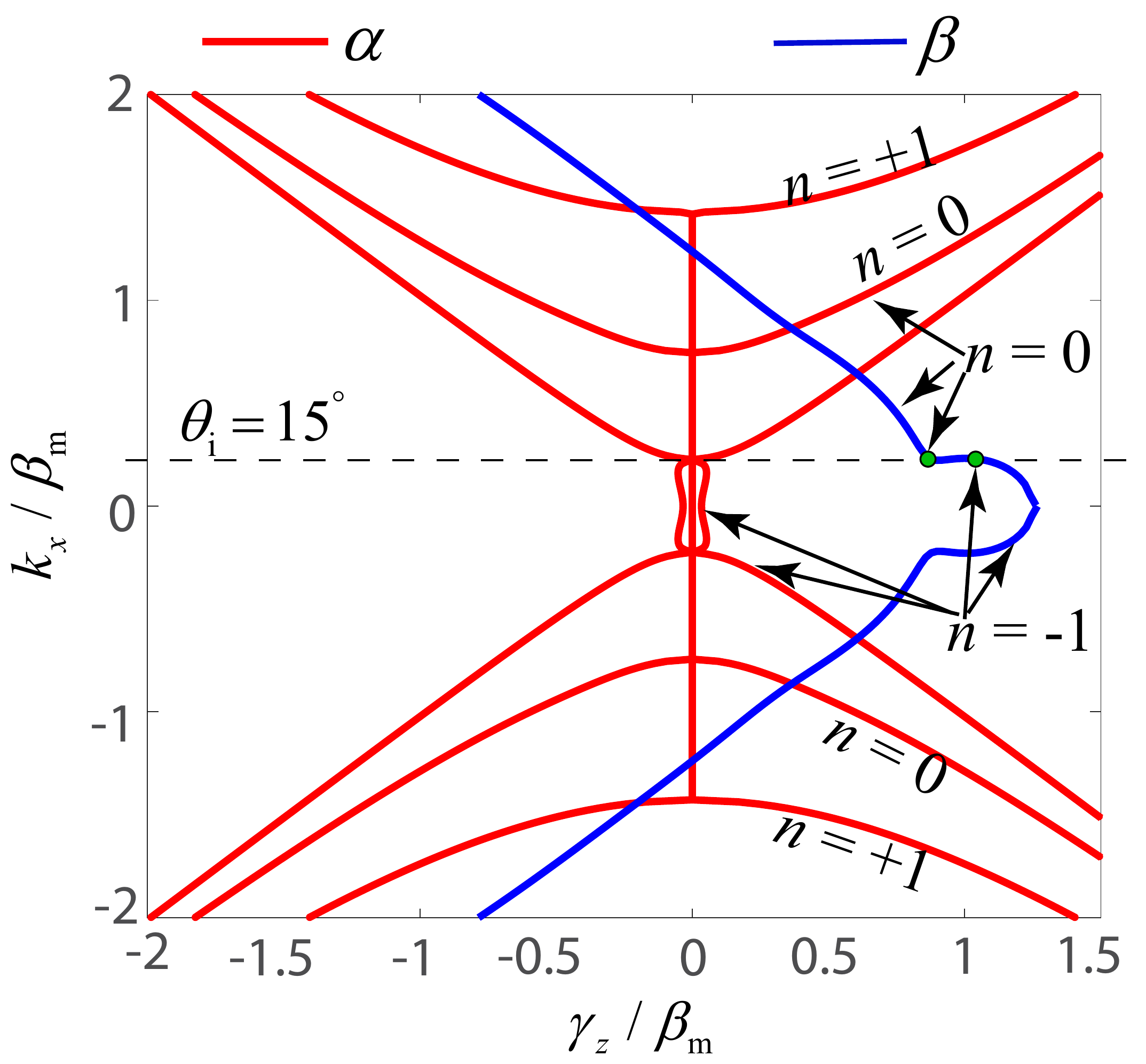}} 	
	\subfigure[]{\label{fig:disp_5}	
		\includegraphics[width=1.1\columnwidth]{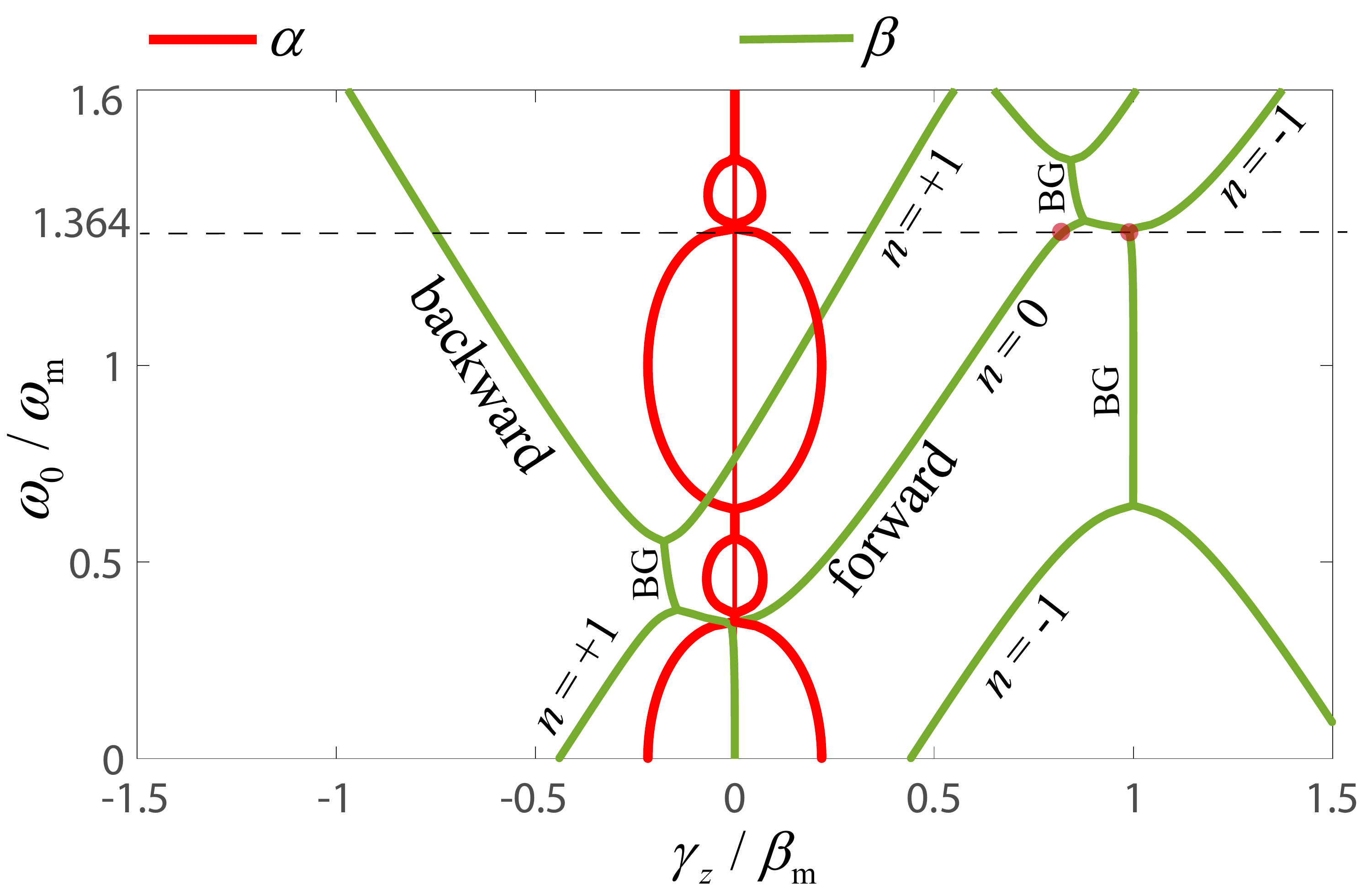}}	
	\caption{Analytical dispersion diagram of the sinusoidally space-time surface wave medium with the electric permittivity in~\eqref{eqa:perm} for subluminal regime, i.e., $\varGamma=0.55$. (a)~Isofrequency diagram, i.e., $\beta_{z,n}(k_x)$ at $\omega/\omega_\text{m}=1.363$, and for $\epsilon_\text{m}\rightarrow 0$. (b)~Same as (a) except for greater modulation amplitude $\epsilon_\text{m}=0.45$. (c)~Isofrequency diagram of the medium in (b), which includes the real and imaginary parts of the $\gamma_{z,n}$, i.e., $\beta_{z,n}$ and $\alpha_{z,n}$, for $n=0$ and $n=-1$. (d) Dispersion diagram, i.e., $\gamma_{z,n}(\omega_{0})$ for the medium in (a) and (b) for the $k_x=0.218$ (corresponding to $\theta_\text{i}=15^\circ$) cut.}
	\label{fig:disper2}
\end{figure*}

Figure~\ref{fig:dispiso_2} plots the analytical isofrequency dispersion diagram ($\beta_{z,n}(k_x)$ at $\omega/\omega_\text{m}=1.363$) of the sinusoidally space-time surface wave medium with the electric permittivity in~\eqref{eqa:perm} for the subluminal regime, i.e., $\varGamma=0.55$ for $\epsilon_\text{m}\rightarrow 0$. It may be seen from this figure that forward $n=0$ and $n=-1$ harmonics are excited very close to each other, where $n=0$ is excited at an angle of scattering of $\theta_{0}=\theta_\text{i}=15^\circ$. However, the $n=-1$ harmonic is intentionally excited at the angle of scattering of $\theta_{-1}=90^\circ$.

Figure~\ref{fig:dispiso_3} plots the same isofrequency $\beta_{z,n}(k_x)$ diagram as Figure~\ref{fig:dispiso_2} except for a greater modulation amplitude of $\epsilon_\text{m}=0.45$. As a result of non-equilibrium in the electric and magnetic permitivitties of the medium, several electromagnetic badgaps appear at the intersections between space-time harmonics~\cite{Taravati_PRAp_2018}. As a consequence, strong coupling between some of the harmonics has occurred, e.g. between the $n=0$ and $n=-1$ harmonics. 

Figure~\ref{fig:dispiso_4} plots the complex isofrequency dispersion diagram $\gamma_{z,n}(k_x)$ for the medium in Fig.~\ref{fig:dispiso_3}. This diagram is formed by two different curves, i.e., the real $\beta_{z,n}(k_x)$ and the imaginary $\alpha_{z,n}(k_x)$ parts of the wavenumber. For the sake of clarification, we have included only a few number of harmonics. This figure shows that at the excited angle of incidence $\theta_\text{i}=15^\circ$, the $n=0$ harmonic introduces a pure real wavenumber, i.e., $\gamma_{z,0}=\beta_{z,0}$, whereas the $n=-1$ harmonic acquires a pure imaginary wavenumber, that is, $\gamma_{z,-1}=i\alpha_{z,-1}$. As a result, a perfect space-time transition from a pure propagating wave to a pure space-time surface wave is ensured.

Figure~\ref{fig:disp_5}	plots the complex dispersion diagram, i.e., $\gamma_{z,n}(\omega_{0})$ for the medium in Fig.~\ref{fig:dispiso_3} and~\ref{fig:dispiso_4}. This figure shows the strong transition between $n=0$ and $n=-1$ harmonics from the $k_x=0.218$ (corresponding to $\theta_\text{i}=15^\circ$) cut.

\subsection{Approximate closed-form solution}\label{sec:cols_sol}

To best investigate the efficiency of the antenna-mixer-amplifier, we present an approximate closed-form solution. Such an approximate closed-form solution clearly shows the effect of modulation parameters on the efficiency of the proposed antenna-mixer-amplifier medium.

\subsubsection{Reception state (down-link)}

Equations (S7) and (S9) in the supplemental material form a matrix differential equation. For the down-link, the initial conditions of $E_{0}(0)=E_\text{i}$ and $E_{-1}(0)=0$ gives
\begin{equation}
E_{0}(z)=E_\text{i}      \cos\left( \dfrac{\epsilon_\text{m} k_0 k_{-1}}{4  \sqrt{\gamma_{z,-1}\gamma_{z,0} }   } z  \right),
\end{equation}
\begin{equation}
E_{-1}(z)=i E_\text{i}  \dfrac{k_{-1}   }{k_0} \sqrt{ \dfrac{\gamma_{z,0}  }{\gamma_{z,-1}} }\sin \left( \dfrac{\epsilon_\text{m} k_0 k_{-1}}{4  \sqrt{\gamma_{z,-1}\gamma_{z,0} }   } z  \right),
\end{equation}
where $k_{-1}=\omega_{-1}/c^2$. The optimal transition from the incident field to the space-time surface wave occurs at $z=d$, where
\begin{equation}
\frac{d}{dz} E_{-1}(z)|_{z=d}=0,
\end{equation}
yielding $|\epsilon_\text{m} k_0 k_{-1} d/ (4  \sqrt{\gamma_{z,-1}\gamma_{z,0} } )|   =\pi/2$
which corresponds to
\begin{align}\label{eq:iii}
d=2\pi  
\left| \dfrac{ \sqrt{\gamma_{z,-1}\gamma_{z,0} } } {\epsilon_\text{m} k_0 k_{-1}}  \right|.
\end{align}

The down-link transition power ratio reads
\begin{equation}
\text{G}_\text{d}=\dfrac{|E_{-1}(z=d)|^2}{|E_\text{i}|^2}=\left|\dfrac{  k_{-1}^2 \gamma_{z,0}}{ k_0^2 \gamma_{z,-1} } \right|=\dfrac{  \omega_{-1}^2 \beta_\text{m}}{ \omega_0^2 \alpha_{z,-1} }, 
\label{eq:gd}    
\end{equation}

Therefore, for a set of modulation parameters, i.e., $\omega_\text{m}$, $\alpha_{z,-1}$ and $\varGamma$, a down-link power gain may be achieved. For the particular case, with the dispersion diagram in Figs.~\ref{fig:dispiso_4} and~\ref{fig:disp_5}, since $\alpha_{z,-1}<<\beta_\text{m}$ (i.e., $\alpha_{z,-1}/\beta_\text{m}=0.03$ and $\omega_{-1}^2/\omega_{0}^2=0.071$), the down-link power gain reads $\text{G}_\text{d}=3.75$~dB. This power gain emerges from the coupling of the space-time modulation power to the incident wave~\cite{Cullen_NAT_1958,tien1958traveling,Taravati_Kishk_PRB_2018,li2019nonreciprocal}.

The total down-link electric field inside the slab reads
\begin{equation}\label{eqa}
\begin{split}
&E(x,z,t)=\\
&E_\text{i} \cos\left( \dfrac{\epsilon_\text{m} k_0 k_{-1}}{4  \sqrt{i\alpha_{z,-1}\beta_{z,0} }   } z  \right) e^{-i \left(k_x x+\beta_{z,0} z -\omega_0 t \right)}+ \\
&
i E_\text{i}  \dfrac{k_{-1}   }{k_0} \sqrt{ \dfrac{\beta_{z,0}  }{i\alpha_{z,-1}} } \sin\left( \dfrac{\epsilon_\text{m} k_0 k_{-1}}{4  \sqrt{i\alpha_{z,-1}\beta_{z,0} }   } z  \right) e^{-i \left(k_x x+i\alpha_{z,-1}z -\omega_{-1} t \right)},
\end{split}
\end{equation}

\subsubsection{Transmission state (up-link)}

For the up-link, the matrix differential equations (S7) and (S9) with the initial conditions of $E_{-1}(0)=E_\text{i}$ and $E_{0}(0)=0$ yields
\begin{equation}
E_{-1}(z)=E_\text{i}      \cos\left( \dfrac{\epsilon_\text{m} k_0 k_{-1}}{4  \sqrt{\gamma_{z,-1}\gamma_{z,0} }   } z  \right),
\end{equation}
\begin{equation}
E_{0}(z)=i E_\text{i}  \dfrac{k_{0}   }{k_{-1}} \sqrt{ \dfrac{\gamma_{z,-1}  }{\gamma_{z,0}} }\sin \left( \dfrac{\epsilon_\text{m} k_0 k_{-1}}{4  \sqrt{\gamma_{z,-1}\gamma_{z,0} }   } z  \right).
\end{equation}

The up-link transition power ratio reads
\begin{equation}
\text{G}_\text{u}=\dfrac{|E_{0}(z=d)|^2}{|E_\text{i}|^2}=\left|\dfrac{  k_{0}^2 \gamma_{z,-1}}{ k_{-1}^2 \gamma_{z,0} } \right|=\dfrac{  \omega_{0}^2 \alpha_{z,-1}}{ \omega_{-1}^2 \beta_\text{m}  }, 
\label{eq:gu}    
\end{equation}
which is the inverse of the down-link transition power ratio. This shows that the proposed antenna-mixer-amplifier introduces a power amplification only in one direction (here down-link), as shown in Fig.~\ref{fig:symb}. The total up-link electric field inside the slab reads
\begin{equation}\label{eqa}
\begin{split}
&E(x,z,t)=\\
&i E_\text{i}  \dfrac{k_{0}   }{k_{-1}} \sqrt{ \dfrac{i\alpha_{z,-1}   }{\beta_{z,0}} } \sin\left( \dfrac{\epsilon_\text{m} k_0 k_{-1}}{4  \sqrt{i\alpha_{z,-1}\beta_{z,0} }   } z  \right) e^{-i \left(k_x x+\beta_{z,0} z -\omega_0 t \right)}+ \\
&E_\text{i} \cos\left( \dfrac{\epsilon_\text{m} k_0 k_{-1}}{4  \sqrt{i\alpha_{z,-1}\beta_{z,0} }   } z  \right) 
e^{-i \left(k_x x+i\alpha_{z,-1}z -\omega_{-1} t \right)}.
\end{split}
\end{equation}

\section{Results}\label{sec:Res}
\begin{figure*}
	\subfigure[]{\label{fig:siw_comp}
		\includegraphics[width=1.4\columnwidth]{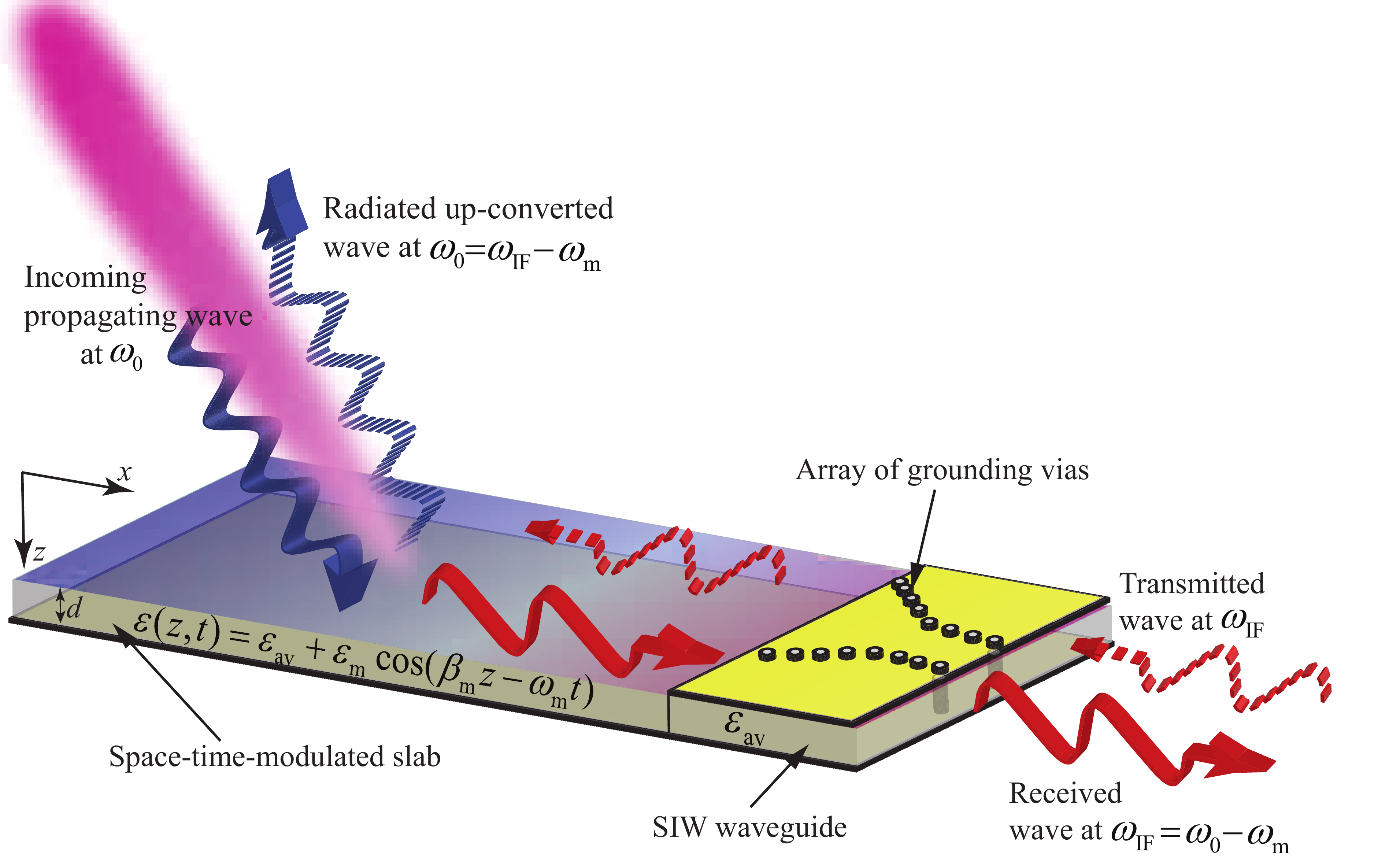}}
	\subfigure[]{\label{fig:FDTD_time_DL}
		\includegraphics[width=1.2\columnwidth]{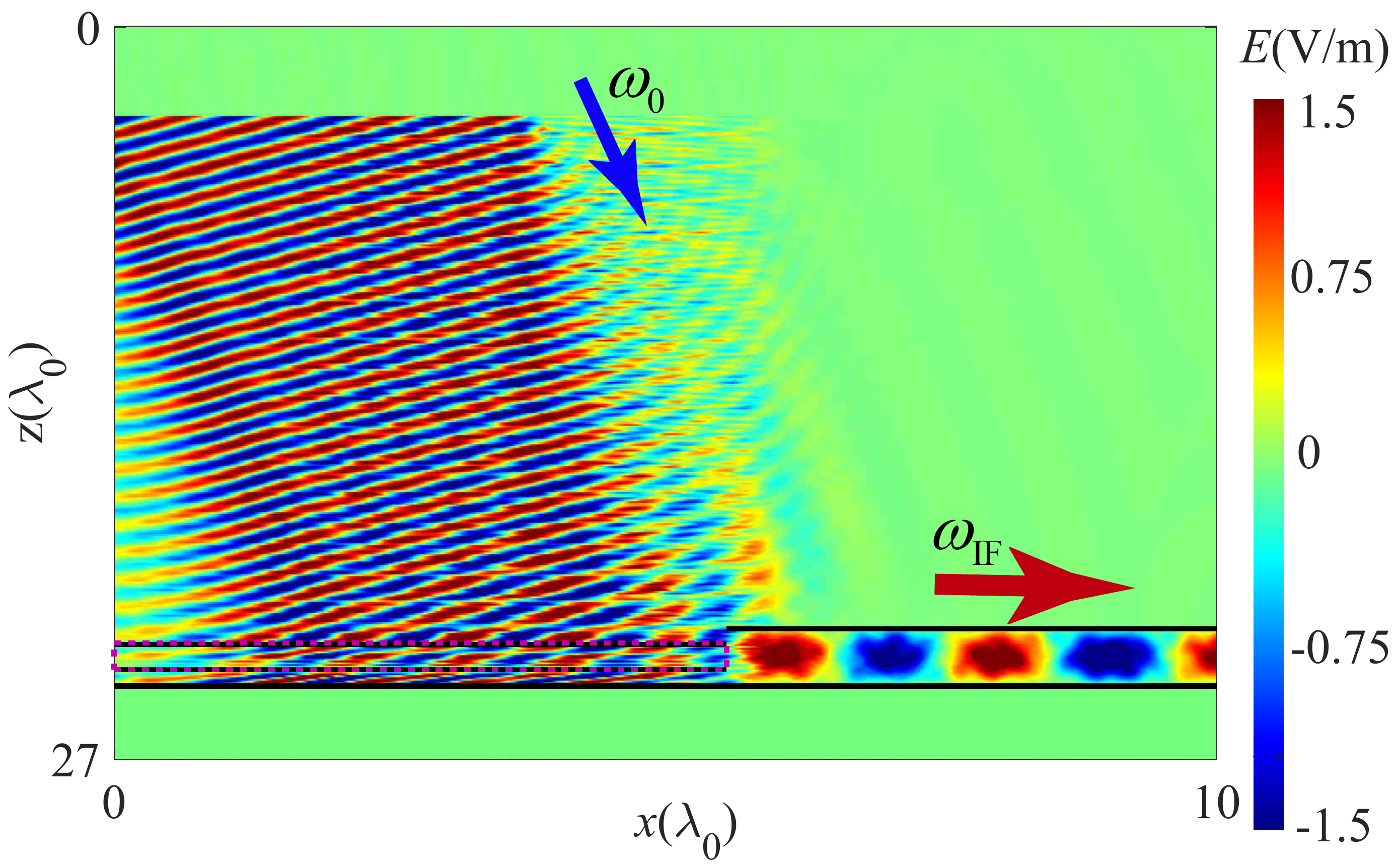}}
	\subfigure[]{\label{fig:FDTD_spec_DL}
		\includegraphics[width=0.8\columnwidth]{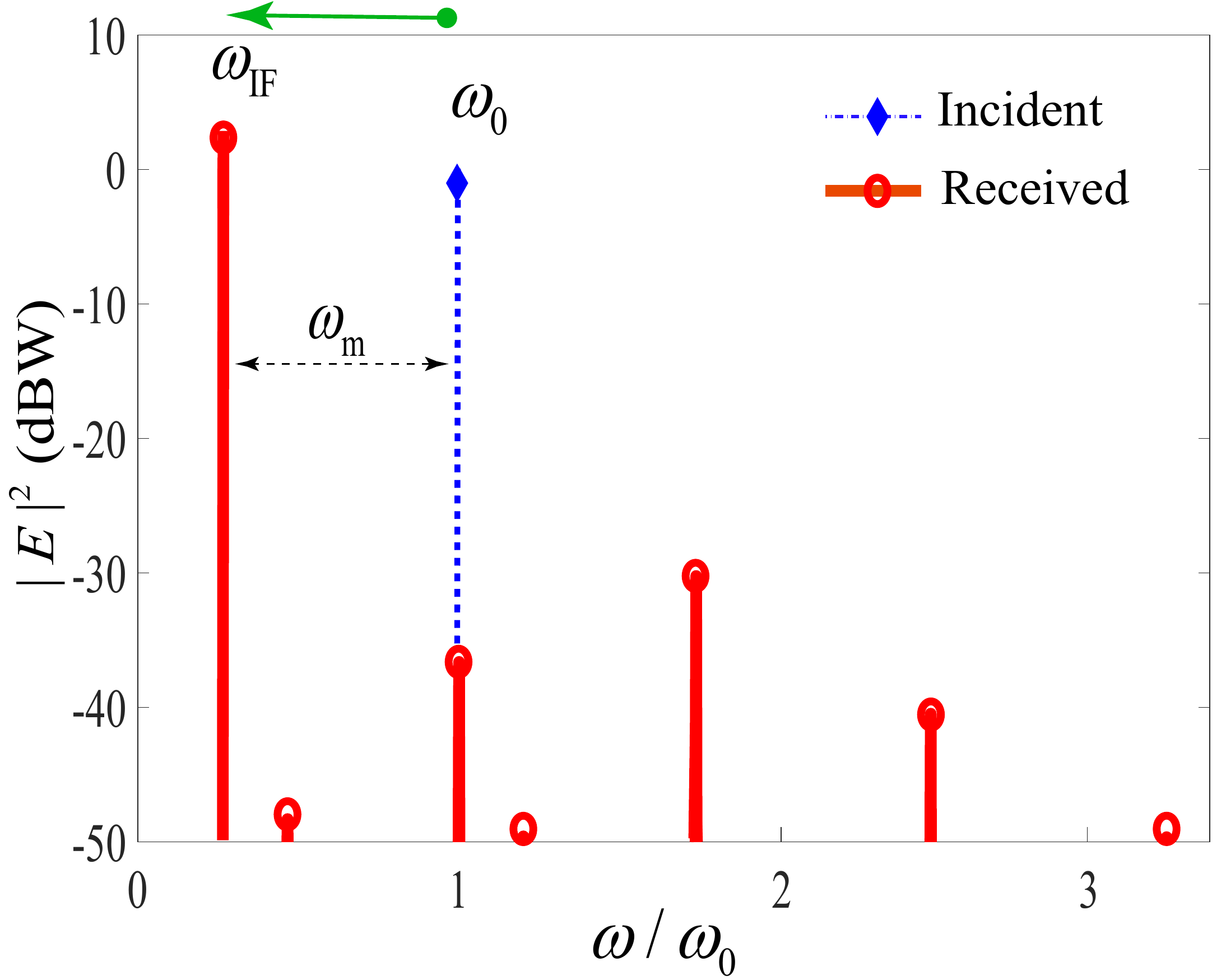}}
	\caption{FDTD simulation results of the antenna-mixer-amplifier space-time surface wave medium with $\omega_0/\omega_\text{m}=1.363$, $\epsilon_\text{m}=0.45$, $d=0.8 \lambda_0$, and $\theta_\text{i}=15^\circ$. (a) Architecture of the medium. (b) Time- and (c) frequency-domain responses for the down-link transition.}
	\label{fig:FDTD}
\end{figure*}

This section investigates the functionality and efficiency of the antenna-mixer-amplifier space-time slab using FDTD numerical simulation. We consider oblique incidence to the slab and compare the numerical results with the analytical solution provided in Secs.~\ref{sec:anal_sol} and~\ref{sec:cols_sol}. Figure~\ref{fig:siw_comp} depicts the structure of the antenna-mixer-amplifier space-time surface wave medium. A substrate integrated waveguide (SIW) is integrated with the space-time slab to efficiently launch and receive space-time surface waves. The slab assumes $\omega_\text{m}=2\pi\times2.2$~GHz, $\epsilon_\text{m}=0.45$, $\varGamma=0.55$, and $d=0.8 \lambda_0$. A plane wave with temporal frequency $\omega_0=2\pi\times3$~GHz is propagating in the $+z$-direction under an angle of incidence of $\theta_\text{i}=15^\circ$, and impinges on the slab.

Figure~\ref{fig:FDTD_time_DL} shows the time domain numerical simulation result for the receiving state (down-link). It may be seen from this figure that a pure transition from the incident space-wave at $\omega_0$ to the space-time surface wave, propagating along the $x$-direction, at frequency $\omega_\text{IF}=\omega_\text{0}-\omega_\text{m}=2\pi\times 0.8$ GHz occurs. Furthermore, it may be seen from Fig.~\ref{fig:FDTD_time_DL} that the amplitude of the received wave is stronger than the amplitude of the incident wave. Figure~\ref{fig:FDTD_spec_DL} plots the frequency domain numerical simulation result. This plot clearly shows a pure and strong transition (frequency-conversion) from the incident wave to the down-converted space-time surface wave. The 3.5 dB power conversion gain is in agreement with the analytical result in Eq.~\eqref{eq:gd}. In addition, the amplitudes of the undesired harmonics are more than $33$ dB lower than the amplitude of the down-converted harmonic at $\omega_\text{IF}$.

\begin{figure*}
	\subfigure[]{\label{fig:FDTD_time_UL}
		\includegraphics[width=1.2\columnwidth]{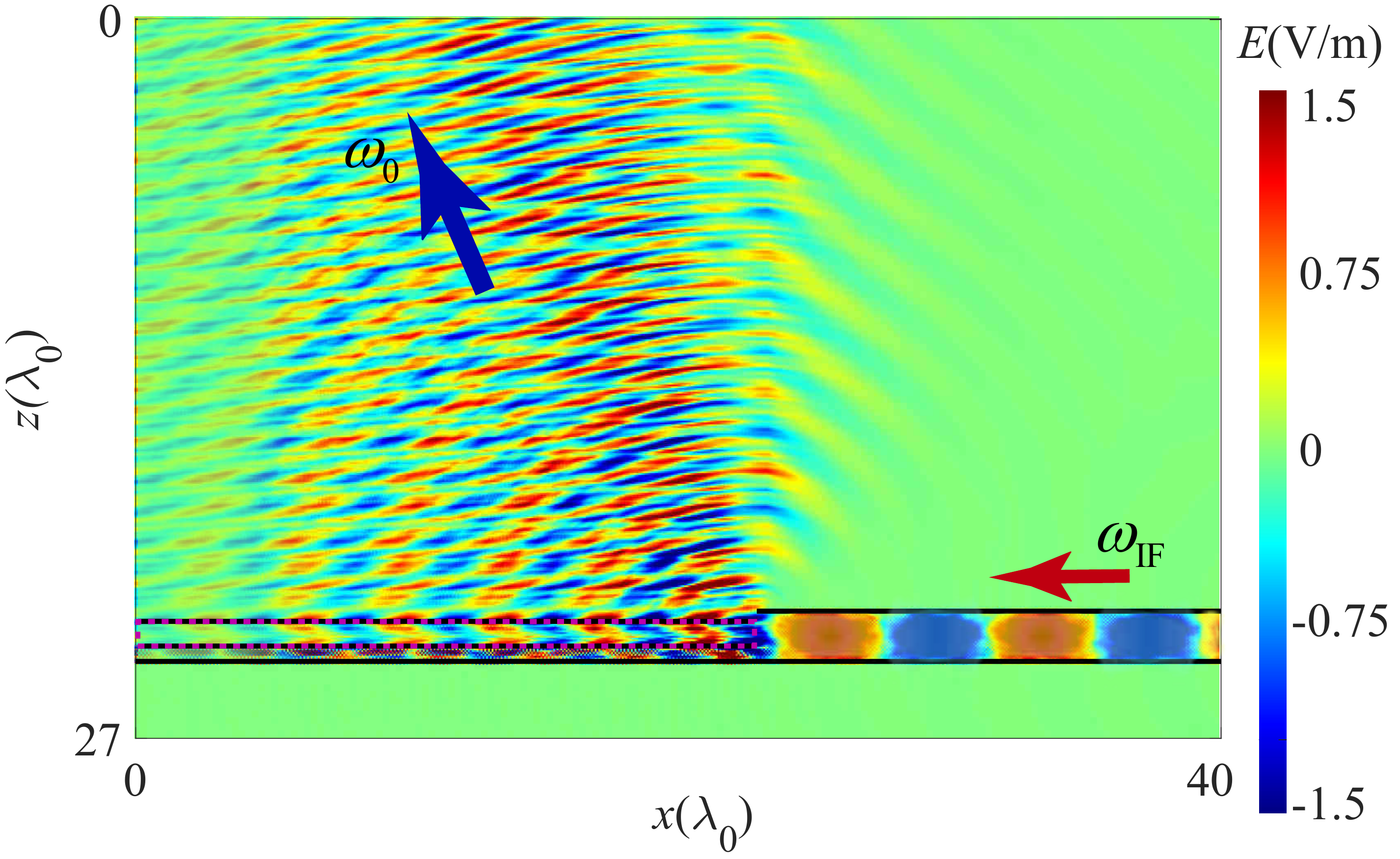}}  
	\subfigure[]{\label{fig:FDTD_spec_UL}
		\includegraphics[width=0.8\columnwidth]{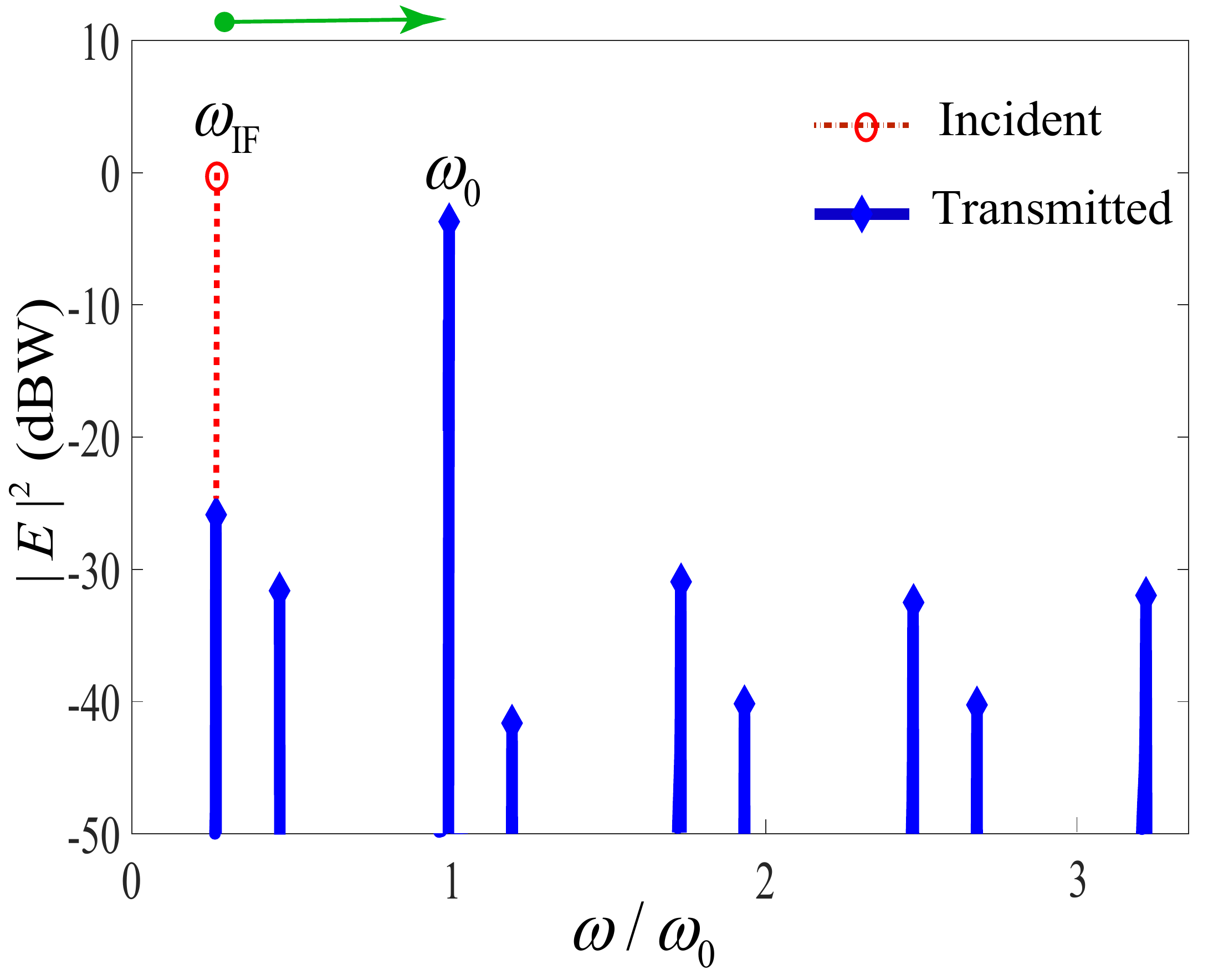}}			
	\caption{FDTD simulation results of the antenna-mixer-amplifier space-time surface wave medium with $\omega_0/\omega_\text{m}=1.363$, $\epsilon_\text{m}=0.45$, $d=0.8 \lambda_0$, and $\theta_\text{i}==15^\circ$. (a) and (b) Time-domain and frequency-domain responses for the up-link transition.}
	\label{fig:FDTD2}
\end{figure*}

Figure~\ref{fig:FDTD_time_UL} shows the time domain FDTD numerical simulation result for the transmission state (up-link). Here, a transition (up-conversion) from the space-time surface wave at $\omega_\text{IF}$ to the space-wave at $\omega_\text{0}=\omega_\text{IF}+\omega_\text{m}$ occurs.

Figure~\ref{fig:FDTD_spec_UL} plots the frequency domain numerical simulation result for the transmission state (up-link). This plot clearly shows a pure and strong transition (frequency-conversion) from the incident wave to the down-converted space-time surface wave. The 3.52 dB power conversion loss is in agreement with the analytical result in Eq.~\eqref{eq:gu}. In addition, the amplitude of the undesired harmonics are more than $27$ dB lower than the amplitude of the down-converted harmonic at $\omega_\text{0}$.

\section{Conclusions}\label{sec:conc}
We showed that a space-time-varying medium operates as an antenna-mixer-amplifier. Such a unique functionality is achieved by taking advantages of space-time surface waves associated with complex space-time wave vectors in a subluminal space-time medium. The proposed structure is endowed with pure frequency up- and down-conversions and very weak undesired time harmonics. In contrast to the recently proposed space-time mixers, a large frequency ratio between the incident wave frequency and the up-/down-converted wave frequency, with a down- or up-conversion gain, is achievable. Furthermore, as the structure does not operate based on progressive energy transition between the space-time modulation and the incident wave, it possesses a subwavelength thickness (metasurface). Such a multi-functional, highly efficient and compact medium represents a new class of integrated electronic-electromagnetic component, and is expected to find various applications in modern wireless telecommunication systems.

\bibliographystyle{IEEEtran}
\bibliography{Taravati_Reference}

\newpage

\begin{widetext}

\qquad\qquad\qquad\qquad\qquad\qquad\qquad \qquad\textbf{{SUPPLEMENTAL MATERIAL}}

\vspace{0.6cm}

	\renewcommand{\theequation}{S\arabic{equation}}
\renewcommand{\thesubsection}{\arabic{subsection}}
\renewcommand{\thesubsubsection}{3.\arabic{subsubsection}}
\setcounter{equation}{0}  
\setcounter{subsection}{0}  

\headsep = 40pt

The antenna-mixer-amplifier space-time medium is placed between $z=0$ and $z=d$, and represented by the space-time-varying permittivity of
\begin{equation}\label{eqa:permit}
\epsilon_\text{eq}(z,t)=\epsilon_\text{av} + \epsilon_\text{m} \cos(\beta_\text{m} z-\omega_\text{m} t),
\end{equation}
and $\mu=\mu_0$. The electric field inside the slab is defined based on the superposition of the $m=0$ and $m=-1$ space-time harmonics fields, i.e.,
\begin{align}\label{eqa:el}
E_\text{S}(x,z,t)=E_{0}(z) e^{-i \left(k_x x+\gamma_{z,0} z -\omega_0 t \right)}+E_{-1}(z) e^{-i \left(k_x x+\gamma_{z,-1}z -\omega_{-1} t \right)},
\end{align}
where $\omega_{-1}=\omega_{0}-\omega_\text{m}$. The corresponding homogeneous wave equation reads
\begin{align}\label{eqa:wave_eq}
\frac{\partial^{2} \textbf{E}}{\partial x^{2}}+\frac{\partial^{2} \textbf{E}}{\partial z^{2}}= \frac{1}{c^2} \frac{\partial^{2} [\epsilon_\text{eq}(t,z) \textbf{E}]}{\partial t^{2}}.
\end{align}

Inserting the electric field in~\eqref{eqa:el} into the wave equation in~\eqref{eqa:wave_eq} results in 
\begin{equation}
\begin{split}
&\left( \frac{\partial^{2} }{\partial x^{2}}+\frac{\partial^{2} }{\partial z^{2}}  \right) \left[  E_{0}(z) e^{-i \left(k_x x+\gamma_{z,0} z -\omega_0 t \right)}+ E_{-1}(z) e^{-i \left(k_x x+\gamma_{z,-1}z -\omega_{-1} t \right)}   \right]\\
& \qquad  = \frac{1}{c^2} \frac{\partial^{2} }{\partial t^{2}} 
\left(\bigg[\epsilon_\text{av} + \frac{\epsilon_\text{m}}{2} e^{i(\beta_\text{m} z-\omega_\text{m} t)}  + \frac{\epsilon_\text{m}}{2} e^{-i(\beta_\text{m} z-\omega_\text{m} t)} \right] \\
& \qquad \qquad  \times \left(E_{0}(z) e^{-i \left(k_x x+\gamma_{z,0} z -\omega_0 t \right)}+ E_{-1}(z) e^{-i \left(k_x x+\gamma_{z,-1}z -\omega_{-1} t \right)} \right) \bigg),
\end{split}
\end{equation}
and applying the space and time derivatives, while using a slowly varying
envelope approximation (i.e., $\partial^{2} E_{-1}(z)/\partial z^{2}=0$ and $\partial^{2} E_{0}(z)/\partial z^{2}=0$), yields
\begin{equation}\label{eqa:eq1}
\begin{split}
& \left[(k_x^2+\gamma_{z,0}^2 )E_{0}(z) -2i\gamma_{z,0} \frac{\partial E_{0}(z)}{\partial z} \right] e^{-i \left(k_x x+\gamma_{z,0} z -\omega_0 t \right)}\\
&\qquad+\left[(k_x^2+\gamma_{z,-1}^2 ) E_{-1}(z) -2i\gamma_{z,-1} \frac{\partial E_{-1}(z)}{\partial z} \right] e^{-i \left(k_x x+\gamma_{z,-1}z -\omega_{-1} t \right)}\\
& \qquad\qquad  = \frac{1}{c^2} 
\left(\bigg[\omega_0^2\epsilon_\text{av}+ \frac{\epsilon_\text{m}}{2} (\omega_0-\omega_\text{m})^2 e^{i(\beta_\text{m} z-\omega_\text{m} t)}  + \frac{\epsilon_\text{m}}{2}(\omega_0+\omega_\text{m})^2 e^{-i(\beta_\text{m} z-\omega_\text{m} t)} \right] E_{0}(z) e^{-i \left(k_x x+\gamma_{z,0} z -\omega_0 t \right)}\\
&\qquad\qquad +\left[\omega_{-1}^2\epsilon_\text{av} + \frac{\epsilon_\text{m}}{2} (\omega_{0}-2\omega_\text{m})^2 e^{i(\beta_\text{m} z-\omega_\text{m} t)}  + \frac{\epsilon_\text{m}}{2}  \omega_0^2 e^{-i(\beta_\text{m} z-\omega_\text{m} t)} \right] E_{-1}(z) e^{-i \left(k_x x+\gamma_{z,-1}z -\omega_{-1} t \right)}  \bigg).
\end{split}
\end{equation}

We then multiply both sides of Eq.~\eqref{eqa:eq1} with $e^{i \left(k_x x+\gamma_{z,0} z -\omega_0 t \right)}$, which gives
\begin{equation}\label{eqa:eq2}
\begin{split}
& \left[(k_x^2+\gamma_{z,0}^2 )E_{0}(z) -2i\gamma_{z,0} \frac{\partial E_{0}(z)}{\partial z} \right] \\
& +\left[(k_x^2+\gamma_{z,-1}^2 ) E_{-1}(z) -2i\gamma_{z,-1} \frac{\partial E_{-1}(z)}{\partial z} \right]  e^{i \left(\beta_\text{m}-\omega_\text{m} t \right)}\\
&  
\qquad\qquad  = \frac{1}{c^2} 
\left(\bigg[\omega_0^2\epsilon_\text{av}+ \frac{\epsilon_\text{m}}{2} (\omega_0-\omega_\text{m})^2 e^{i(\beta_\text{m} z-\omega_\text{m} t)}  + \frac{\epsilon_\text{m}}{2}(\omega_0+\omega_\text{m})^2 e^{-i(\beta_\text{m} z-\omega_\text{m} t)} \right] E_{0}(z) \\
&\qquad\qquad \qquad +\left[\omega_{-1}^2\epsilon_\text{av} + \frac{\epsilon_\text{m}}{2} (\omega_{0}-2\omega_\text{m})^2 e^{i(\beta_\text{m} z-\omega_\text{m} t)}  + \frac{\epsilon_\text{m}}{2}  \omega_0^2 e^{-i(\beta_\text{m} z-\omega_\text{m} t)} \right] E_{-1}(z) e^{i \left(\beta_\text{m}-\omega_\text{m} t \right)}  \bigg),
\end{split}
\end{equation}
and next, applying $\int_{0}^{\frac{2\pi}{\omega_\text{m}}}dt$ to both sides of~\eqref{eqa:eq2} yields

\begin{equation}\label{eqa:eqcoup1}
\frac{d E_{0}(z)}{d z}=  i \dfrac{\epsilon_\text{m} k_0^2}{4  \gamma_{z,0} }  
E_{-1}(z).
\end{equation}

Following the same procedure, we next multiply both sides of~\eqref{eqa:eq2} with $e^{-i \left(\beta_\text{m} z -\omega_\text{m} t \right)}$, and applying $\int_{0}^{\frac{2\pi}{\omega_\text{m}}}dt$ in both sides of the resultant, which reduces to
\begin{equation}
\left[(k_x^2+\gamma_{z,-1}^2 ) E_{-1}(z) -2i\gamma_{z,-1} \frac{d E_{-1}(z)}{d z} \right] 
= \frac{\omega_{-1}^2}{c^2} 
\left( \frac{\epsilon_\text{m}}{2}    E_{0}(z)  +\epsilon_\text{av}  E_{-1}(z)  \right),
\end{equation}

\begin{equation}\label{eqa:eqcoup2}
\frac{d E_{-1}(z)}{d z} 
= i\dfrac{\epsilon_\text{m} k_{-1}^2}{4  \gamma_{z,-1} }    E_{0}(z),
\end{equation}
where $k_{-1}=\omega_{-1}/c^2$. Solving the coupled equations~\eqref{eqa:eqcoup1} and~\eqref{eqa:eqcoup2} together, we achieve the field coefficients $E_0(z)$ and $E_{-1}(z)$ as given in Eqs.(17) and (18) for the reception state and in Eqs.(23) and (24) for the transmission state.

\end{widetext}
\end{document}